\documentclass[prd,showpacs]{revtex4}
\begin{document}

\title{Supersymmetric non conservative systems}

\author{N.E. Mart\'{\i}nez-P\'erez}
 \email{nephtalieliceo@hotmail.com}
  \affiliation{Benem\'erita Universidad Aut\'onoma de Puebla, Facultad de Ciencias
F\'{\i}sico Matem\'ati vbcas, P.O. Box 165, 72000 Puebla, M\'exico.}
 
\author{C. Ram\'{\i}rez}
 \email{cramirez@fcfm.buap.mx}
 \affiliation{Benem\'erita Universidad Aut\'onoma de Puebla, Facultad de Ciencias
F\'{\i}sico Matem\'aticas, P.O. Box 165, 72000 Puebla, M\'exico.}

 \date{\today}

\begin{abstract}
We give the generalization of a recent variational formulation for nonconservative classical mechanics, for fermionic and sypersymmetric systems. Both cases require slightly modified boundary conditions. The supersymmetric version is given in the superfield formalism. The corresponding Noether theorem is formulated. As expected, like the energy, the supersymmetric charges are not conserved.
\end{abstract}
\pacs{11.30.Pb,12.60.Jv}
\maketitle

\section{Introduction}
The study of a general mechanical system includes influences of external factors, whose origin and detailed description may be partially or fully unknown. The evolution of such systems is frequently irreversible and non invariant under time reversal. There are fundamental questions related to this issue, like time direction, the second principle of thermodynamics, etc. A formulation for such systems has been given in terms of equations of motion. For instance, for a nonconservative subsystem of a conservative system, we can obtain a set of equations of motion by means of the substitution of the solution of the  equations of the rest of the system into the ones of the subsystem. This reasoning is extended to write ``effective'' equations of motion of nonconservative systems.

Hamilton's principle gives a way to obtain the equations of motion of conservative systems by the variation of the action, with the variables fixed at the initial and final times. The physical trajectory goes along a curve through these initial and final points and minimizes the action. Thus this trajectory is determinated by conditions in the past and in the future. For conservative, time reversible, systems, this situation all right. However if one wishes to describe nonconservative systems, not invariant under time reversal, such conditions could lead to contradictory situations.

Classical mechanics deals with the evolution of systems described by real quantities like the position, angles, and so on. However, if we see it as a limit of quantum mechanics, it is natural to ask about fermionic degrees of freedom, which have properties properly described by anticommutative quantities. This has lead to the formulation of fermionic classical mechanics \cite{casalbuoni}, whose quantization leads to the quantum mechanics of fermionic systems \cite{teitelboim}. Supersymmetry transforms bosons into fermions and its basic entities are supermultiplets, which contain bosons and fermions, see e.g. \cite{wess} and \cite{bellucci}. The profound difference between bosonic and fermionic systems has been one of the principal reasons which made supersymmetry so attractive. On the other side, supersymmetry imposes strong restrictions, like the cancellation of divergences in field theory, one of the main reasons of its success, or the equality of the masses of particles in the same supermultiplet, one of the principal objections against it as a symmetry of nature. 

In mechanics, supersymmetry can be realized by an extension of time translations, by means of transformations with anticommutative parameters. Thus supersymmetry is intimately related to energy conservation and a question which arises naturally is what happens with non-closed, nonconservative systems. On the other side, if a supersymmetric system is exposed to external forces, these forces could be due to supersymmetric unknown factors. Thus, such a theory could have a supersymmetric structure. 

In two recent papers, \cite{galley1,galley2}, a systematic proposal has been made for a lagrangian formulation for nonconservative systems. This proposal is similar to the closed time path formalism, and amounts to a modification of Hamilton's variational principle, introducing a generalized nonconservative ``potential". The closed time path formalism has been originally proposed by Schwinger \cite{schwinger} for field theory, for further development see e.g. \cite{chou}. It has been developed recently for classical and quantum mechanics in \cite{polonyi}.

In this paper we follow the formulation of \cite{galley1,galley2}, to address the study of supersymmetric nonconservative systems. We consider first fermionic systems, whose boundary conditions are determined by only one parameter \citep{teitelboim}. Thus their nonconservative generalization requires a slight modification of the boundary conditions. The generalization for supersymmetric systems is done in the superspace formalism. The boundary conditions must be also modified, and are given in terms of superfields.  Considering that supersymmetry in classical mechanics is an extension of time translational symmetry, and that the last is broken by the nonconservative interaction, the same must happen for supersymmetry, as is verified by means of the Noether theorem. We maintained the supersymmetric structure for the nonconservative generalized potential, i.e it is written in terms of superfields. 
In Section II we give a short review of the formulation of \cite{galley1}. In section III we consider the generalization for fermionic systems. In Section IV we perform the supersymmetric generalization and formulate the corresponding Noether theorem. We consider along the work the examples of two coupled oscillators and a damped oscillator. In the last Section we draw some conclusions.

\section{Lagrangian Formulation}

Hamilton's principle in mechanics establishes that the trajectory of a system in an arbitrary time interval $[t_{i},t_{f}]$, on which act conservative forces, is an extremum of the action. The variation is done on curves which go through two fixed points, at the initial and final times. Instead of it, we could consider curves that at the initial time go through a fixed point, with a fixed velocity. In this case the point at the final time should be fixed also, although it cannot be arbitrary, as it is determined by the initial position and velocity. It can be seen that in this way we obtain the Euler-Lagrange equations as well \cite{arnold}. As far as the kinetic and potential terms do not depend explicitly on time, the E-L equations are symmetric under time reversal, although the integration constants of the solutions can be chosen asymmetric. However, if we are interested on an effective action, where a part of the degrees of freedom have been integrated, by substituting their solutions into the action, the resulting action will be as well time reversal invariant. For instance \cite{galley1}, for a system of two coupled oscillators, after integrating one of them, the equation of motion of the resulting effective lagrangian has a reversible interaction, with a time symmetric Green's function. Thus, in general, in this way the equations of motion of an irreversible, nonconservative, system cannot be obtained.

The lagrangian formulation of Galley supposes that there is a conservative lagrangian $L(q,\dot{q})$, where $q$ are $n$-dimensional vectors. Nonconservativity is attained by a doubling of the degrees of freedom $q\rightarrow(q_1,q_2)$, and the variational principle is modified in such a way that $q_{2}(t)$ effectively runs back in time, and appears as a continuation of $q_{1}(t)$. Thus the boundary conditions for the variation are that, at the initial time both variables are fixed, and at the final time their values coincide, but have an arbitrary variation. Once the variation done, the doubling is reverted, at the level of the equations of motion, by the so called physical limit $q_1\rightarrow q_2=q$. This is similar to the Closed Path-Time approach \cite{schwinger,chou}. In fact, both variables could be arranged as one, beginning and finishing at $t_i$, after a closed time path $t_i\rightarrow t_f\rightarrow t_i$ \cite{polonyi}. 
Galley avoids the time loop and instead of it modifies the setting of the action
\begin{equation}
S=\int_{t_i}^{t_f}L({q}_1,{\dot{q}}_1)dt+\int_{t_f}^{t_i}L({q}_2,{\dot{q}}_2)dt=\int_{t_i}^{t_f}L({q}_1,{\dot{q}}_1)dt-\int_{t_i}^{t_f}L({q}_2,{\dot{q}}_2)dt.
\end{equation}
This setting allows to add to the action a nonconservative ``generalized potential", $K(q_1,\dot q_1,q_2,\dot q_2)$, which depends on both variables and is antisymmetric under $1\leftrightarrow 2$ 
\begin{equation}
K(q_1,\dot q_1,q_2,\dot q_2)=-K(q_2,\dot q_2,q_1,\dot q_1).\label{antisym}
\end{equation}
Thus, te nonconservative lagrangian is given by
\begin{equation}
\Lambda(q_1,\dot q_1,q_2,\dot q_2)=L({q}_1,{\dot{q}}_1)-L({q}_2,{\dot{q}}_2)+K(q_1,\dot q_1,q_2,\dot q_2).
\end{equation}
The variation is given now with the boundary conditions that at the initial time $\delta{q}_1(t_i)=\delta{q}_2(t_i)=0$, and at the final time both variables coincide ${q}_1(t_f)={q}_2(t_f)$, and have arbitrary variations $\delta{q}_1(t_f)=\delta{q}_2(t_f)$. Aditionally, ${\dot {q}}_1(t_f)={\dot {q}}_2(t_f)$. Thus
\begin{eqnarray}
\delta S&=&\int_{t_{i}}^{t_{f}}\delta\Lambda(q_1,\dot q_1,q_2,\dot q_2)dt
=\left[\delta q_{1}\left(\frac{\partial L}{\partial{\dot q}_{1}}+\frac{\partial K}{\partial{\dot q}_{1}}\right)+\delta q_{2}\left(-\frac{\partial L}{\partial{\dot q}_{2}}+\frac{\partial K}{\partial{\dot q}_{2}}\right)\right]_{t=t_{f}}\nonumber\\
&+&\int_{t_{i}}^{t_{f}}\left\{\delta q_{1}\left[\frac{\partial}{\partial q_{1}}(L+K)-\frac{d}{dt}\frac{\partial}{\partial{\dot q}_{1}}(L+K)\right]-\delta q_{2}\left[\frac{\partial}{\partial q_{2}}(L-K)-\frac{d}{dt}\frac{\partial}{\partial{\dot q}_{2}}(L-K)\right]\right\}dt,
\end{eqnarray}
where summation indices are understood. The boundary terms vanish after taking into account the boundary conditions and the antisymmetry of $K$, from which in particular follows
\begin{equation}
\left.\left(\frac{\partial K}{\partial{\dot q}_{1}}+\frac{\partial K}{\partial{\dot q}_{2}}\right)\right|_{q_1=q_2,\ \dot q_1=\dot q_2}=0.\label{derk}
\end{equation}
Thus, the equations of motion are
\begin{eqnarray}
\left(\frac{\partial}{\partial q_{1}}-\frac{d}{dt}\frac{\partial}{\partial{\dot q}_{1}}\right)\left[L(q_1,\dot{q_1})+K(q_1,\dot{q_1},q_2,\dot{q_2})\right]=0,\label{eq1}\\
\left(\frac{\partial}{\partial q_{2}}-\frac{d}{dt}\frac{\partial}{\partial{\dot q}_{2}}\right)\left[L(q_2,\dot{q_2})-K(q_1,\dot{q_1},q_2,\dot{q_2})\right]=0.\label{eq2}
\end{eqnarray}
Due to (\ref{antisym}), in the physical limit  $q_{1}=q_{2}=q$, the second equation is identical with the first one. Thus, the nonconservative equations of motion are
\begin{eqnarray}
\frac{\partial L}{\partial q}-\frac{d}{dt}\frac{\partial L}{\partial{\dot q}}
=F_{K}\equiv\left.\left(\frac{\partial}{\partial q_{2}}-\frac{d}{dt}\frac{\partial}{\partial{\dot q}_{2}}\right)K(q_{1},\dot{q}_{1},q_{2},\dot{q}_{2})\right\vert_{q_{1}=q_{2}=q}.
\end{eqnarray}

\subsection{Coupled oscillators}
This formulation allows to treat coupled oscillators in a reversing time asymmetric way \cite{galley1}. Consider the lagrangian
\begin{equation}
L(q,\dot q,Q,\dot Q)=\frac{m}{2}\left(\dot{q}^2-\omega^2 q^2\right)+\lambda qQ+
\frac{M}{2}\left(\dot{Q}^2-\omega^2 Q^2\right).\label{cosc}
\end{equation}
The two equations of motion of this action, can be solved by the substitution of the solution of one of the equations, into the other. This can be done at the level of the action (\ref{cosc}), substituting therein the solution for $Q$, i.e. integrating it out. In this way an effective action is obtained. The causal solution of the equation of motion of $Q$ is
\begin{equation}
Q(t)=Q^{(h)}(t)+\frac{\lambda}{M}\int_{t_i}^{t_f}G_{\rm ret}(t-t')q(t')dt'\label{qsol}
\end{equation}
where $Q^{(h)}$ is the homogeneous solution. Substituting (\ref{qsol}) into the equation of motion of $q$, we get
\begin{equation}
\ddot q(t)+m\omega^2 q(t)-\lambda Q^{(h)}(t)-\frac{\lambda^2}{M}\int_{t_i}^{t_f}G_{\rm ret}(t-t')q(t')dt'.\label{eqq}
\end{equation}
Otherwise, the substitution of (\ref{qsol}) into (\ref{cosc}), leads to the ``effective'' action
\begin{equation}
S_{\rm eff}=\int_{t_i}^{t_f}dt\left[\frac{m}{2}\left(\dot{q}^2-\omega^2 q^2\right)+\frac{\lambda}{2}qQ^{(h)}\right]+
\frac{\lambda^2}{2M}\int_{t_i}^{t_f}dt\int_{t_i}^{t_f}dt'\ q(t)G_{\rm ret}(t-t')q(t')+{\rm B.T.},\label{cosce}
\end{equation}
whose equation of motion is not causal because it contains only the symmetric part of the Green function $G_{\rm ret}$. The boundary terms in (\ref{cosce}), due to the partial integration $\int \dot{q}^2 dt\rightarrow -\int q\ddot q dt+q(t_{f})\dot q(t_{f})-q(t_{i})\dot q(t_{i})$, have a nonvanishing variation, unless $\delta \dot q(t_i)=\delta \dot q(t_f)=0$, as required for actions with second order derivatives $\ddot q$. Thus, if the variation is done as usual for first order actions, the equations of motion will contain an arbitrary parameter. In fact these boundary terms could be discarded, but they should be kept for a general formulation, consistent with path integral.

The corresponding nonconservative lagrangian is $\Lambda=L(q_1,\dot q_1,Q_1,\dot Q_1)-L(q_2,\dot q_2,Q_2,\dot Q_2)$, and can be written more conveniently in the coordinates $q_{\pm}=1/{\sqrt 2}(q_1\pm q_2)$, $Q_{\pm}=1/{\sqrt 2}(Q_1\pm Q_2)$, which satisfy $\delta q_{\pm}(t_{i})=\delta Q_{\pm}(t_{i})=0$, $q_{-}(t_{f})=Q_{-}(t_{f})=0$ and $\dot q_{-}(t_{f})=\dot Q_{-}(t_{f})=0$.
\begin{equation}
\Lambda=m\left(\dot q_{+}\dot q_{-}-\omega^2q_{+}q_{-}\right)+\lambda\left(q_{+}Q_{-}+q_{-}Q_{+}\right)+M\left(\dot Q_{+}\dot Q_{-}-\omega^2Q_{+}Q_{-}\right).\label{coscpm}
\end{equation}
The causal solutions of the equations of motion of $Q_{\pm}$, can be substituted into this action in order to obtain the effective action. However, as in the case of (\ref{cosce}), there will be boundary terms whose variation do not vanish. This problem can be avoided if we note that there is an ambiguity due to $\int \dot Q_{+}\dot Q_{-}dt=-\int(aQ_{+}\ddot Q_{-}+bQ_{-}\ddot Q_{+})dt-aQ_{+}(t_{i})\dot Q_{-}(t_{i})-bQ_{-}(t_{i})\dot Q_{+}(t_{i})$, $a+b=1$. Thus the vanishing of the variation of the boundary terms requires $\delta\dot Q_{\pm}(t_i)=0$. Otherwise, if the boundary terms are discarded, setting $a=0$, the effective action is
\begin{equation}
S_{\rm eff}=\int_{t_i}^{t_f}dt\left[m\left(\dot{q}_{+}\dot{q}_{-}-\omega^2 q_{+}q_{-}\right)+\lambda q_{-}Q^{(h)}_{+}\right]+
\frac{\lambda^2}{M}\int_{t_i}^{t_f}dt\int_{t_i}^{t_f}dt'\ q_{-}(t)G_{\rm ret}(t-t')q_{+}(t'),\label{coscpme}
\end{equation}
whose equation of motion in the physical limit is the causal one (\ref{eqq}).

\subsection{Damped oscillator}
Another example is the damped oscillator \cite{galley1}. The conservative lagrangian is the one of the oscillator $L(q,\dot q)=m/2(\dot q^{2}-\omega^{2}q^{2})$ and $K(q_{1},\dot q_{1},q_{2},\dot q_{2})=c/2(q_{1}\dot q_{2}-q_{2}\dot q_{1})$. Hence
\begin{eqnarray}
\Lambda(q_1,q_2,\dot q_1,\dot q_2)=\frac{m}{2}(\dot q_1^{2}-\omega^{2}q_1^{2})-\frac{m}{2}(\dot q_2^{2}-\omega^{2}q_2^{2})+\frac{c}{2}(q_{1}\dot q_{2}-q_{2}\dot q_{1}).
\end{eqnarray}
The equation of motion, in the physical limit is
\begin{equation}
m\ddot q-c\dot q+m\omega^2q=0.
\end{equation}
\section{Fermionic systems}
Fermionic systems in classical mechanics are formulated by means of anticommuting variables. We consider variables $\psi$ and $\bar\psi$, where $\bar\psi$ is the complex conjugated. These variables can have an $n$-dimensional index and satisfy $\psi_{i}\psi_{j}=-\psi_{j}\psi_{i}$, $\psi_{i}\bar\psi_{j}=-\bar\psi_{j}\psi_{i}$, and $\bar\psi_{i}\bar\psi_{j}=-\bar\psi_{j}\bar\psi_{i}$, otherwise they commute with bosonic quantities. Thus they are nilpotent, $\psi_{i}\psi_{i}=0$. Complex conjugation reverses the order like hermitian conjugation, thus giving a sign, e.g. $(\psi_{i}\psi_{j})^{\dagger}=-\bar\psi_{i}\bar\psi_{j}$. In the following we will not indicate the $n$-dimensional indices. We will adopt the convention that fermionic derivatives act on the left, as 
\begin{equation}
\frac{\partial}{\partial\psi_i} \psi_j=\delta_{ij}-\psi_j\frac{\partial}{\partial\psi_i}.
\end{equation}

The kinetic term of fermionic systems is first order, which means that the trajectory is determined by fixing only one parameter. For this reason, in \cite{teitelboim} a boundary term has been proposed for fermionic actions. Consider for example a system described by one fermionic variable $\psi(t)$ with lagrangian $L(\psi,\bar\psi)=i/2(\bar\psi\dot\psi+\psi\dot{\bar\psi})+\psi\bar\psi$. We use an economical notation, where the dependence of the Lagrangian on the first time derivatives is not indicated. According to \cite{teitelboim}, to the action are added suitable boundary terms
\begin{equation}
S=\int_{t_{i}}^{t_{f}}\left[\frac{i}{2}(\bar\psi\dot\psi+\psi\dot{\bar\psi})-\omega\psi\bar\psi\right]dt-
\frac{i}{2}\left[\psi(t_{1})\bar\psi(t_{2})+\bar\psi(t_{1})\psi(t_{2})\right],
\end{equation}
whose variation with the boundary conditions $\delta\psi(t_{1})+\delta\psi(t_{2})=0$ and $\delta\bar\psi(t_{1})+\delta\bar\psi(t_{2})=0$, gives the expected equations of motion $i\dot\psi+\omega\psi=0$ and $i\dot{\bar\psi}-\omega\bar\psi=0$. However, this procedure depends on the action and seems to be incompatible with supersymmetry. 

Instead of adding boundary terms, we proceed following the observations at the beginning of the previous section. Thus we fix the values of the fermionic coordinates at the initial and final times, but with the consistency restriction that one of both values is determined by the other. As long as Hamilton's principle requires only that the variations of the coordinates at the initial and final times vanish, there is no problem if we do not bother about how is this dependence, which in fact turns out after solving the equations of motion.
On the other side, a nonconservative formulation for fermions, after the doubling of the variables $\psi\rightarrow(\psi_{1},\psi_{2})$ and $\bar\psi\rightarrow(\bar\psi_{1},\bar\psi_{2})$, requires that $\psi_{1}(t_{f})=\psi_{2}(t_{f})$, $\bar\psi_{1}(t_{f})=\bar\psi_{2}(t_{f})$. This means that for consistency, also the values at the initial time should coincide $\psi_{1}(t_{i})=\psi_{2}(t_{i})$, $\bar\psi_{1}(t_{i})=\bar\psi_{2}(t_{i})$. Furthermore, a general form of the nonconservative potential require the conditions $\dot\psi_{1}(t_{f})=\dot\psi_{2}(t_{f})$ and $\dot{\bar{\psi}}_{1}(t_{f})=\dot{\bar{\psi}}_{2}(t_{f})$.
Therefore the action is
\begin{equation}
S=\int_{t_{i}}^{t_{f}}\left[L(\psi_{1},\bar\psi_{1})-L(\psi_{2},\bar\psi_{2})+K(\psi_{1},\psi_{2},\bar\psi_{1},\bar\psi_{2})\right]dt,
\end{equation}
with $K(\psi_{1},\psi_{2},\bar\psi_{1},\bar\psi_{2})$ antisymmetric under $1\leftrightarrow 2$. The boundary conditions are $\delta\psi_{a}(t_{i})=\delta\bar\psi_{a}(t_{i})=0$ $(a=1,2)$, $\psi_{1}(t_{i,f})=\psi_{2}(t_{i,f})$, $\bar\psi_{1}(t_{i,f})=\bar\psi_{2}(t_{i,f})$, $\delta\psi_{1}(t_{f})=\delta\psi_{2}(t_{f})$ and $\delta\bar\psi_{1}(t_{f})=\delta\bar\psi_{2}(t_{f})$. Following the same steps as in the preceding section and considering the relations corresponding to (\ref{derk}) in the present case,
we get the same equations of motion as in (\ref{eq1}) and (\ref{eq2})
\begin{eqnarray}
\frac{\partial}{\partial \psi_{1}}(L+K)-\frac{d}{dt}\frac{\partial}{\partial{\dot\psi}_{1}}(L+K)=0,\label{eqf1}\\
\frac{\partial}{\partial \psi_{2}}(L-K)-\frac{d}{dt}\frac{\partial}{\partial{\dot\psi}_{2}}(L-K)=0,\label{eqf2}
\end{eqnarray}
and their complex conjugated $\psi\rightarrow\bar\psi$. Again, these equations coincide in the physical limit $\psi_{1}=\psi_{2}=\psi$ and $\bar\psi_{1}=\bar\psi_{2}=\bar\psi$. Hence the nonconservative equations of motion are
\begin{eqnarray}
\frac{\partial L}{\partial \psi}-\frac{d}{dt}\frac{\partial L}{\partial{\dot\psi}}
&=&F_{K}\equiv\left.\left(\frac{\partial}{\partial\psi_{2}}-\frac{d}{dt}\frac{\partial}{\partial{\dot\psi}_{2}}\right)K(\psi_{1},\bar{\psi}_{1},\dot{\psi}_{1},\dot{\bar{\psi}}_{1},\psi_{2},\dot{\psi}_{2},\bar{\psi}_{2},\dot{\bar{\psi}}_{2})\right\vert_{\psi_{1}=\psi_{2}=\psi},\\
\frac{\partial L}{\partial \bar{\psi}}-\frac{d}{dt}\frac{\partial L}{\partial{\dot{\bar{\psi}}}}
&=&\bar F_{K}\equiv\left.\left(\frac{\partial}{\partial\bar{\psi}_{2}}-\frac{d}{dt}\frac{\partial}{\partial{\dot{\bar{\psi}}}_{2}}\right)K(\psi_{1},\bar{\psi}_{1},\dot{\psi}_{1},\dot{\bar{\psi}}_{1},\psi_{2},\dot{\psi}_{2},\bar{\psi}_{2},\dot{\bar{\psi}}_{2})\right\vert_{\psi_{1}=\psi_{2}=\psi}.
\end{eqnarray} 

\subsection{Coupled fermionic oscilators}
As an example, let us consider two coupled fermionic oscillators with Lagrangian
\begin{equation}
L(\psi,\bar\psi,\Psi,\bar\Psi)=-i\frac{m}{2}\left(\bar\psi\dot\psi+\psi\dot{\bar\psi}\right)+m\omega\psi\bar\psi-\lambda\left(\psi\bar\Psi+\Psi\bar\psi\right)-
i\frac{M}{2}\left(\bar\Psi\dot\Psi+\Psi\dot{\bar\Psi}\right)+M\Omega\Psi\bar\Psi.
\end{equation}
Similar to the bosonic case (see e.g. \cite{galley1}), the effective action obtained by means of Hamilton's principle by the integration of $\Psi$ and $\bar\Psi$ is given by
\begin{eqnarray}
S_{{\rm eff}}&=&\int_{t_{i}}^{t_{f}}\left[-i\frac{m}{2}\left(\bar\psi\dot\psi+\psi\dot{\bar\psi}\right)+m\omega\psi\bar\psi-\frac{\lambda}{2}\left(\psi{\bar\Psi}^{(h)}+\Psi^{(h)}\bar\psi\right)\right]\nonumber\\
&-&\frac{\lambda^2}{2M}\int_{t_{i}}^{t_{f}}dt\int_{t_{i}}^{t_{f}}dt'\psi(t)\left[\bar\Gamma_{ret}(t-t')+\bar\Gamma_{adv}(t-t')\right]\bar\psi(t'),
\label{lfint}
\end{eqnarray}
where $\Psi^{(h)}=e^{i\Omega t}\Psi_{0}$ is the solution of the homogeneous equation, and the retarded and advanced fermionic Green functions are
\begin{eqnarray}
\bar\Gamma_{ret}(t-t')&=&-\frac{i}{2}\,\theta(t-t')e^{-i\Omega(t-t')},\label{gret}\\
\bar\Gamma_{adv}(t-t')&=&\frac{i}{2}\,\theta(t'-t)e^{-i\Omega(t-t')}.\label{gadv}
\end{eqnarray} 
This effective action is time reversal symmetric. In order to obtain an irreversible effective action, the nonconservative fermionic formulation can be used, with a Lagrangian $\Lambda=L(\psi_1,\bar\psi_1,\Psi_1,\bar\Psi_1)-L(\psi_2,\bar\psi_2,\Psi_2,\bar\Psi_2)$. If we use the notation $\psi_{\pm}=1/\sqrt{2}(\psi_1\pm\psi_2)$, this Lagrangian can be written as
\begin{eqnarray}
\Lambda=-i\frac{m}{2}\left(\psi_{+}{\dot{\bar\psi}}_{-}+{\bar\psi}_{-}{\dot{\psi}}_{+}\right)+m\omega\psi_{+}{\bar\psi}_{-}-\lambda\left(\psi_{+}{\bar\Psi}_{-}+\psi_{-}{\bar\Psi}_{+}\right)
-i\frac{M}{2}\left(\Psi_{+}{\dot{\bar\Psi}}_{-}+{\bar\Psi}_{-}{\dot{\Psi}}_{+}\right)+M\Omega\Psi_{+}{\bar\Psi}_{-}+{\rm c.c.},\label{accionf}
\end{eqnarray}
where c.c. means complex conjugated. The boundary conditions with these variables are $\delta\psi_{+}(t_i)=\delta\psi_{-}(t_i)=\delta\Psi_{+}(t_i)=\delta\Psi_{-}(t_i)=0$, $\delta\psi_{-}(t_f)=\delta\Psi_{-}(t_f)=0$, $\psi_{-}(t_{i,f})=\Psi_{-}(t_{i,f})=0$, plus their complex conjugated.
Thus, the equations of motion of $\Psi$ and $\bar\Psi$ are
\begin{eqnarray}
i\dot\Psi_{\pm}-\Omega\Psi_{\pm}+\frac{\lambda}{M}\psi_{\pm}=0,\label{ecf1}\\
i{\dot{\bar\Psi}}_{\pm}+\Omega\bar\Psi_{\pm}-\frac{\lambda}{M}\bar\psi_{\pm}=0,\label{ecf2}
\end{eqnarray}
with the following causal solutions
\begin{eqnarray}
\Psi_{\pm}(t)&=&\Psi_\pm^{(h)}(t)-\frac{\lambda}{M}\int_{t_{i}}^{t_{f}}dt'\Gamma_{ret}(t-t')\psi_\pm(t'),\label{solf1}\\
\bar\Psi_{\pm}(t)&=&\bar\Psi_\pm^{(h)}(t)-\frac{\lambda}{M}\int_{t_{i}}^{t_{f}}dt'\bar\Gamma_{ret}(t-t')\bar\psi_\pm(t'),\label{solf2}
\end{eqnarray}
where $\Psi_{-}^{(h)}(t)=\bar\Psi_{-}^{(h)}(t)=0$ due to the boundary conditions. The substitution of these solutions to obtain the effective action is ambiguous as in the bosonic case. Indeed, including the conjugated terms, action (\ref{accionf}) can be written as 
\begin{eqnarray}
\Lambda=-i\frac{m}{2}\left(\psi_{+}{\dot{\bar\psi}}_{-}+{\bar\psi}_{-}{\dot{\psi}}_{+}\right)+m\omega\psi_{+}{\bar\psi}_{-}-\lambda\left(\psi_{+}{\bar\Psi}_{-}+\psi_{-}{\bar\Psi}_{+}\right)
-iM\left(a\Psi_{+}{\dot{\bar\Psi}}_{-}+b\bar\Psi_{-}{\dot{\Psi}}_{+}\right)
+M\Omega\Psi_{+}{\bar\Psi}_{-}
+{\rm c.c.},\label{accionfa}
\end{eqnarray}
where $a+b=1$. The boundary terms $iM(a-1/2)[\Psi_{+}(t_f)\bar\Psi_{-}(t_f)-\Psi_{+}(t_i)\bar\Psi_{-}(t_i)]+$c.c. vanish due to the boundary conditions. Thus, after taking into account the equations of motion of $\Psi_{\pm}$ and $\bar\Psi_{\pm}$ in (\ref{accionfa}), we get
\begin{equation}
\Lambda=-i\frac{m}{2}\left(\psi_{+}{\dot{\bar\psi}}_{-}+{\bar\psi}_{-}{\dot{\psi}}_{+}\right)+m\omega\psi_{-}{\bar\psi}_{+}-\lambda\left(a\psi_{+}{\bar\Psi}_{-}-b{\bar\psi}_{-}\Psi_{+}\right)+{\rm c.c.},
\end{equation}
where now $\Psi_{\pm}$ and $\bar\Psi_{\pm}$ are sources, given by (\ref{solf1}) and (\ref{solf2}). Substituting these solutions into the action, we get the effective action
\begin{eqnarray}
S_{{\rm eff}}&=&\int_{t_{i}}^{t_{f}}dt\left[-i\frac{m}{2}\left(\psi_{+}{\dot{\bar\psi}}_{-}+{\bar\psi}_{-}{\dot{\psi}}_{+}\right)+m\omega\psi_{-}{\bar\psi}_{+}+b\lambda{\bar\psi}_{-}\Psi_{+}^{(h)}\right]\nonumber\\
&&+\frac{\lambda^2}{M}\int_{t_{i}}^{t_{f}}dt\int_{t_{i}}^{t_{f}}dt'{\bar\psi}_{-}(t)\left[a\Gamma_{adv}(t-t')+b\Gamma_{ret}(t-t')\right]\psi_{+}(t')+{\rm c.c.},\label{accionf1}
\end{eqnarray}
whose equations of motion in the physical limit, $\psi_{+}=\sqrt{2}\psi$, $\psi_{-}=0$, are
\begin{eqnarray}
i\dot\psi+\omega\psi-\frac{\lambda}{M}\Psi^{(h)}-\frac{\lambda^2}{M^2}\int_{t_{i}}^{t_{f}}dt'\left[a\Gamma_{adv}(t-t')+b\Gamma_{ret}(t-t')\right]{\psi}(t')=0,\nonumber\\
i\dot{\bar\psi}-\omega\bar\psi+\frac{\lambda}{M}\Psi^{(h)}+\frac{\lambda^2}{M^2}\int_{t_{i}}^{t_{f}}dt'\left[b\bar\Gamma_{ret}(t-t')+a\bar\Gamma_{adv}(t-t')\right]{\bar\psi}(t')=0.
\end{eqnarray}
Therefore the solution can be chosen to be causal by setting $a=0$.
\subsection{Damped fermionic oscillator}
Let us now consider a damped fermionic oscillator. For the lagrangian we consider the fermionic oscillator 
\begin{equation}
L(\psi,\bar\psi)=-i\frac{m}{2}\left(\bar\psi\dot\psi+\psi\dot{\bar\psi}\right)+m\omega\psi\bar\psi
\end{equation}
and 
\begin{equation}
K(\psi_{1},\bar\psi_{1},\psi_{2},\bar\psi_{2})=\frac{\tilde c}{2}\left(\psi_{1}\dot{\bar\psi}_{2}-\psi_{2}\dot{\bar\psi}_{1}\right)+{\rm c.c.}.
\end{equation}
Hence $\Lambda=L(\psi_{1},\bar\psi_{1})-L(\psi_{2},\bar\psi_{2})+K(\psi_{1},\bar\psi_{1},\psi_{2},\bar\psi_{2})$, which, after some partial integrations, can be rewritten as
\begin{eqnarray}
\Lambda&=&-i\frac{m}{2}\left(\psi_{+}{\dot{\bar\psi}}_{-}+{\bar\psi}_{-}{\dot{\psi}}_{+}\right)+m\omega\psi_{+}{\bar\psi}_{-}
+\frac{\tilde c}{2}\left(\psi_{-}\dot{\bar\psi}_{+}-\psi_{+}\dot{\bar\psi}_{-}\right)
+{\rm c.c.}\nonumber\\
&=&-im\left(1-i\frac{\tilde c}{m}\right){\bar\psi}_{-}{\dot{\psi}}_{+}-im\left(1+i\frac{\tilde c}{m}\right){\psi}_{-}{\dot{\bar\psi}}_{+}+m\omega\left(\psi_{-}{\bar\psi}_{+}-{\bar\psi}_{-}\psi_{+}\right).
\label{accionfd}
\end{eqnarray}
The equations of motion in the physical limit are 
\begin{eqnarray}
i\dot{\psi}+\tilde\omega\left(1+i\frac{\tilde c}{m}\right)\psi&=&0, \label{eqfnc1}\\
i\dot{\bar{\psi}}-\tilde\omega\left(1-i\frac{\tilde c}{m}\right)\bar\psi&=&0, \label{eqfnc2}
\end{eqnarray}
where $\tilde\omega=\omega/(1+\tilde c^{2}/m^2)$, with solutions
\begin{eqnarray}
\psi(t)=e^{i\tilde\omega(1+i\frac{\tilde c}{m})t}\psi_{0}\qquad{\rm and}\qquad
\bar\psi(t)=e^{-i\tilde\omega(1-i\frac{\tilde c}{m})t}\bar\psi_{0}.
\end{eqnarray}

The energy change rate can be computed as in \cite{galley2}, by means of the Noether theorem, which for a conservative system of fermionic variables works in the same way as for bosonic variables.
\section{Supersymmetric formulation}
Supersymmetric mechanics can be realized by an extension of time to a Grassmann space, or superspace, $t\rightarrow z\equiv(t,\theta,\bar\theta)$, where $\theta$ and $\bar{\theta}$ are anticommuting variables. There are derivatives defined in these spaces by the rules $\{\partial_\theta,\theta\}=1$, $\{\partial_{\bar\theta},\bar\theta\}=1$, $\{\partial_\theta,\bar\theta\}=0$ and $\{\partial_{\bar\theta},\theta\}=0$, and integration
$\int d\theta=0$, $\int d\theta\theta=1$, $\int d\bar\theta=0$, $\int d\bar\theta\bar\theta=-1$. The dynamical variables are extended to superfields $q(t)\rightarrow \phi(t,\theta,\bar\theta)$, which are real and are given by a finite expansion in the fermionic variables as 
\begin{equation}
\phi(t,\theta,\bar{\theta})=q(t)+\theta\psi(t)-\bar{\theta}\bar{\psi}(t)+\theta\bar{\theta}b(t),\label{superf}
\end{equation}
where the variables $q(t), \psi(t), \bar\psi(t)$ and $b(t)$ are called the components of the superfield. Note that
$\theta\bar{\theta}$ is real. The transformations of supersymmetry are generated by the fermionic charges $Q=\frac{d}{d\theta}-i\bar{\theta}\frac{d}{dt}$
and $\bar{Q}=-\frac{d}{d\bar{\theta}}+i\theta\frac{d}{dt}$, which satisfy
$\{Q,\bar{Q}\}\equiv Q\bar Q+\bar QQ=2i\frac{d}{dt}$. Thus a supersymmetric transformation is written as $\delta_\xi\phi(t,\theta,\bar\theta)=(\xi Q-\bar\xi\bar Q)\phi(t,\theta,\bar\theta)=\xi\psi-\bar\xi\bar\psi+\theta\bar\xi(b+i\dot q)-\bar\theta\xi(b-i\dot q)+i\theta\bar\theta(\xi\dot\psi+\bar{\xi}\dot{\bar{\psi}})$, from which, by comparison of the components, turn out the infinitesimal transformations
\begin{equation}
\delta_\xi q=\xi\psi-\bar\xi\bar\psi, \quad\delta_\xi\psi=\bar\xi(b+i\dot q), \quad\delta_\xi\bar\psi=\xi(b-i\dot q) \quad{\rm and}\quad \delta_\xi b=i(\xi\dot\psi+\bar{\xi}\dot{\bar{\psi}}),\label{susyt}
\end{equation}
or in finite form
\begin{equation}
\phi'(t,\theta,\bar\theta)=\phi(t-i\xi\bar\theta-i\bar\xi\theta,\theta+\xi,\bar\theta+\bar\xi).
\end{equation}
Thus supersymmetry transformations can be seen as a certain type of superspace translation.
There are covariant derivatives $D=\frac{d}{d\theta}+i\bar{\theta}\frac{d}{dt}$ and $\bar{D}=-\frac{d}{d\bar{\theta}}-i\theta\frac{d}{dt}$, which satisfy $\{D,\bar{D}\}=-2i\frac{d}{dt}$, $\{Q,D\}=\{Q,\bar{D}\}=\{\bar Q,D\}=\{\bar Q,\bar D\}=0$. Thus, quantities obtained from superfields by the action of covariant derivatives will transform as superfields. The integral of the covariant derivatives of a superfield 
\begin{eqnarray}
D\phi(t,\theta,\bar\theta)&=&\psi+\bar\theta(i\dot{q}-b)-i\theta\bar{\theta}\dot{\psi},\label{covd}\\
\bar D\phi(t,\theta,\bar\theta)&=&\bar\psi-\theta(i\dot{q}+b)+i\theta\bar{\theta}\dot{\bar{\psi}},\label{covbd}
\end{eqnarray}
give
\begin{eqnarray}
\int_{t_{i}}^{t_{f}}dt\int d\theta d\bar\theta D\phi(t,\theta,\bar\theta)&=&-i\int_{t_{i}}^{t_{f}}dt\dot\psi=-i\psi\big\vert_{t_i}^{t_f}=-iD\phi(t,0,0)\big\vert_{t_i}^{t_f},\label{dertotal1}\\
\int_{t_{i}}^{t_{f}}dt\int d\theta d\bar\theta \bar D\phi(t,\theta,\bar\theta)&=&i\int_{t_{i}}^{t_{f}}dt\dot{\bar\psi}=i\bar\psi\big\vert_{t_i}^{t_f}=i\bar D\phi(t,0,0)\big\vert_{t_i}^{t_f}.\label{dertotal2}
\end{eqnarray}
Hence such terms can be added to the Lagrangian without changing the equations of motion. Moreover, integration by parts can be done $\int dt d\theta d\bar\theta\,\phi_1 D\phi_2=-\int dt d\theta d\bar\theta\, D\phi_1 \phi_2+$ boundary terms, where the boundary terms, $-i(q_1\psi_2+q_2\psi_1)\vert_{t_i}^{t_f}$, can be neglected.

There are also chiral superfields which are complex
and satisfy the covariant condition $\bar{D}\phi=0$, and which can be written as $\phi(\tilde t,\theta)=A(\tilde t)+\theta\psi(\tilde t)$, where $\tilde t=t+i\theta\bar\theta$.

The superfield formalism allows to write supersymmetric actions as superspace integrals of superfield Lagrangians, taking advantage that the supersymmetry transformation of the superspace integral of a superfield is $\int d\theta d\bar\theta \delta_{\xi}\phi(t,\theta,\bar\theta)=\int d\theta d\bar\theta (\xi Q-\bar\xi\bar Q)\phi(t,\theta,\bar\theta)$, and that, similar to (\ref{dertotal1}) and (\ref{dertotal2}), 
\begin{eqnarray}
\int_{t_{i}}^{t_{f}}dt\int d\theta d\bar\theta Q\phi(t,\theta,\bar\theta)&=&i\psi\vert_{t_i}^{t_f},\label{q1}\\
\int_{t_{i}}^{t_{f}}dt\int d\theta d\bar\theta \bar Q\phi(t,\theta,\bar\theta)&=&-i\bar\psi\vert_{t_i}^{t_f}.\label{q2}
\end{eqnarray}
Therefore, a lagrangian will be a function of superfields and their covariant derivatives of first order, and the corresponding action will be of the general form
\begin{equation}
S=\int_{t_{i}}^{t_{f}}dt\int d\theta d\bar\theta L(\phi,D\phi,\bar D\phi),\label{accionsusy}
\end{equation}
whose supersymmetry transformation is
\begin{equation}
\delta_\xi S=\int d\theta d\bar\theta\, \delta_{\xi}L(\phi,D\phi,\bar D\phi)=\int d\theta d\bar\theta (\xi Q-\bar\xi\bar Q)L=\left.i\left(\xi L_\theta-i\bar\xi L_{\bar{\theta}}\right)\right\vert_{t_i}^{t_f},\label{trafos}
\end{equation}
where $L_\theta$ and $L_{\bar{\theta}}$ are the $\theta$ and $\bar{\theta}$ components of the expansion (\ref{superf}) of the superfield $L$.

The variation of the action (\ref{accionsusy}) is
\begin{eqnarray}
\delta S&=&\int_{t_{i}}^{t_{f}}dt\int d\theta d\bar\theta\left(\delta\phi\frac{\partial L}{\partial\phi}+\delta D\phi\frac{\partial L}{\partial D\phi}+\delta \bar D\phi\frac{\partial L}{\partial \bar D\phi}\right)\nonumber\\&=&\int_{t_{i}}^{t_{f}}dt\int d\theta d\bar\theta\delta\phi\left(\frac{\partial L}{\partial\phi}- D\frac{\partial L}{\partial D\phi}-\bar D\frac{\partial L}{\partial \bar D\phi}\right)+{\rm B.T.},
\end{eqnarray}
where the boundary terms are
\begin{eqnarray}
{\rm B.T.}&=&\int_{t_{i}}^{t_{f}}dt\int d\theta d\bar\theta\left[D\left(\delta\phi\frac{\partial L}{\partial D\phi}\right)+\bar D\left(\delta\phi\frac{\partial L}{\partial \bar D\phi}\right)\right]\nonumber\\
&=&\left.\left\{\delta q\left[\left(\frac{\partial L}{\partial D\phi}\right)_{\theta}+\left(\frac{\partial L}{\partial \bar D\phi}\right)_{\bar\theta}\right]+\delta\psi\left(\frac{\partial L}{\partial D\phi}\right)_q+\delta\bar\psi\left(\frac{\partial L}{\partial \bar D\phi}\right)_q\right\}\right\vert_{t_{i}}^{t_{f}}.
\end{eqnarray}
Therefore, it is enough if $\delta q(t_{i,f})=0$ and $\delta \psi(t_{i,f})=\delta\bar\psi(t_{i,f})=0$, which for the superfield formulation can be completed to $\delta\phi(t_i,\theta,\bar\theta)=\delta\phi(t_f,\theta,\bar\theta)=0$.
Therefore the equations of motion are
\begin{equation}
\frac{\partial L}{\partial\phi}- D\frac{\partial L}{\partial D\phi}-\bar D\frac{\partial L}{\partial \bar D\phi}=0.\label{ems}
\end{equation}
From these equations, the equations of the components are obtained from the $\theta$-expansion. These equations can be also obtained directly from (\ref{accionsusy}) written in components, i.e. after integrating the fermionic superspace variables.
\subsection{Nonconservative formulation}\label{ncsusy}
As mentioned in the introduction, we are interested on the on supersymmetric systems under the influence of nonconservative forces. We assume that these forces have a supersymmetric structure. Such systems can be described by the superfield formulation, following the lines of the bosonic formalism. As a first step, the variables are duplicated $ \phi(t,\theta,\bar\theta)\rightarrow( \phi_1(t,\theta,\bar\theta), \phi_2(t,\theta,\bar\theta))$. Thus the noncommutative action is
\begin{equation}
S=\int_{t_{i}}^{t_{f}}dt\int d\theta d\bar\theta\left[L(\phi_1,D\phi_1,\bar D\phi_1)-L(\phi_2,D\phi_2,\bar D\phi_2)+K(\phi_1,D\phi_1,\bar D\phi_1,\phi_2,D\phi_2,\bar D\phi_2)\right],\label{accionnc}
\end{equation}
where $K(\phi_1,D\phi_1,\bar D\phi_1,\phi_2,D\phi_2,\bar D\phi_2)$ is antisymmetric under the exchange $1\leftrightarrow 2$.
The variation of this action is
\begin{eqnarray}
\delta S=\int_{t_{i}}^{t_{f}}dt\int d\theta d\bar\theta
&\Bigg\{&\delta\phi_{1}\left[\frac{\partial(L+K)}{\partial\phi_1}-D\frac{\partial(L+K)}{\partial D\phi_1}-\bar{D}\frac{\partial(L+K)}{\partial \bar{D}\phi_1}\right]\nonumber\\
&-&\delta\phi_{2}\left[\frac{\partial(L-K)}{\partial\phi_2}-D\frac{\partial(L-K)}{\partial D\phi_2}-\bar{D}\frac{\partial(L-K)}{\partial \bar{D}\phi_2}\right]\Bigg\}+{\rm B.T.}
\end{eqnarray}
The boundary terms are
\begin{eqnarray}
{\rm B.T.}=\int_{t_{i}}^{t_{f}}dt\int d\theta d\bar\theta
&\Bigg\{&D\left[\delta\phi_{1}\frac{\partial(L+K)}{\partial D\phi_1}\right]+\bar{D}\left[\delta\phi_{1}\frac{\partial(L+K)}{\partial \bar{D}\phi_1}\right]\nonumber\\
&-&D\left[\delta\phi_{2}\frac{\partial(L-K)}{\partial D\phi_2}\right]-\bar{D}\left[\delta\phi_{2}\frac{\partial(L-K)}{\partial \bar{D}\phi_2}\right]\Bigg\},
\end{eqnarray}
which, taking into account (\ref{dertotal1}) and (\ref{dertotal2}), vanish with the boundary conditions $\delta\phi_{1}(t_i,\theta,\bar\theta)=\delta\phi_{2}(t_i,\theta,\bar\theta)=0$, $\delta\phi_{1}(t_f,\theta,\bar\theta)=\delta\phi_{2}(t_f,\theta,\bar\theta)$, $\phi_1(t_f,\theta,\bar\theta)=\phi_2(t_f,\theta,\bar\theta)$, $D\phi_1(t_f,\theta,\bar\theta)=D\phi_2(t_f,\theta,\bar\theta)$ and $\bar{D}\phi_1(t_f,\theta,\bar\theta)=\bar{D}\phi_2(t_f,\theta,\bar\theta)$. Additionally, for consistency, the equality conditions of the fermionic variables at the initial time of the previous section, require the superfield conditions $\phi_1(t_i,\theta,\bar\theta)=\phi_2(t_i,\theta,\bar\theta)$. This can be seen considering that, for a supersymetric invariant theory, if one of the components of two superfields coincide, then both superfields must coincide, as can be seen from (\ref{susyt}). In components, the conditions for the superfields correspond to conditions for the components. Thus for the covariant derivatives, from (\ref{covd}) and (\ref{covbd}), we get $\dot Q_{1}(t_{f})=\dot Q_{2}(t_{f})$, $\dot\psi_{1}(t_{f})=\dot\psi_{2}(t_{f})$ and $\dot{\bar\psi}_{1}(t_{f})=\dot{\bar\psi}_{2}(t_{f})$.

Thus the equations of motion are
\begin{eqnarray}
\left(\frac{\partial}{\partial\phi_1}-D\frac{\partial}{\partial D\phi_1}-\bar{D}\frac{\partial}{\partial \bar{D}\phi_1}\right)\left[L(\phi_1)+K(\phi_1,\phi_2)\right]&=&0,\\
\left(\frac{\partial}{\partial\phi_2}-D\frac{\partial}{\partial D\phi_2}-\bar{D}\frac{\partial}{\partial \bar{D}\phi_2}\right)\left[L(\phi_2)-K(\phi_1,\phi_2)\right]&=&0.
\end{eqnarray}
which coincide similarly to (\ref{eq1}) and (\ref{eq2}) and in the physical limit, $ \phi_1(t,\theta,\bar\theta)= \phi_2(t,\theta,\bar\theta)= \phi(t,\theta,\bar\theta)$. Further, if we define the supersymmetric nonconservative forces\begin{equation}
F_K(\phi,D\phi,\bar{D}\phi)=\left.\left(D\frac{\partial}{\partial D\phi_1}+\bar{D}\frac{\partial}{\partial \bar{D}\phi_1}-\frac{\partial}{\partial\phi_1}\right)K(\phi_1,\phi_2)\right\vert_{\phi_1=\phi_2=\phi},
\end{equation}
then, the nonconservative equations of motion are
\begin{equation}
\left(\frac{\partial}{\partial\phi}-D\frac{\partial}{\partial D\phi}-\bar{D}\frac{\partial}{\partial \bar{D}\phi}\right)L(\phi,D\phi,\bar{D}\phi)=F_K(\phi,D\phi,\bar{D}\phi).\label{eqsusy}
\end{equation}
\subsection{Noether theorem}
Before we turn to examples, let us work out, following \cite{galley2}, Noether's theorem for a general supersymmetry transformation, which includes time translations. Thus the transformations are 
\begin{eqnarray}
t\rightarrow t'&=&t+\delta t-i\xi\bar\theta-i\bar{\xi}\theta,\label{trafot}\\
\phi'(t',\theta',\bar{\theta}')&=&\phi(t,\theta,\bar{\theta})+\delta_s\phi(t,\theta,\bar{\theta})=\phi(t,\theta,\bar{\theta})+\delta t\dot{\phi}(t,\theta,\bar{\theta})+(\xi Q-\bar{\xi}\bar{Q})\phi(t,\theta,\bar{\theta}).\label{trafof}
\end{eqnarray}
The transformation of the action is
\begin{eqnarray}
\delta_s S&=&\int_{t_{i}}^{t_{f}}dt\int d\theta d\bar\theta\Bigg\{
\delta t\left[\dot{\phi}\frac{\partial L}{\partial\phi}+\frac{d}{dt}(D\phi)\frac{\partial L}{\partial D\phi}+\frac{d}{dt}(\bar{D}\phi)\frac{\partial L}{\partial \bar{D}\phi}-\frac{d L}{dt}+\frac{\partial L}{\partial t}\right]\nonumber\\&&
+\xi\left[(Q\phi)\frac{\partial L}{\partial\phi}+(QD\phi)\frac{\partial L}{\partial D\phi}+(Q\bar{D}\phi)\frac{\partial L}{\partial \bar{D}\phi}\right]-\bar\xi\left[(\bar Q\phi)\frac{\partial L}{\partial\phi}+(\bar QD\phi)\frac{\partial L}{\partial D\phi}+(\bar Q\bar{D}\phi)\frac{\partial L}{\partial \bar{D}\phi}\right]\Bigg\},
\end{eqnarray}
which, after some arrangements and considering that the supersymmetry transformations anticommute with the covariant derivatives, turns to
\begin{eqnarray}
\delta_s S&=&\int_{t_{i}}^{t_{f}}dt\int d\theta d\bar\theta\Bigg\{
\delta_s\phi\left(\frac{\partial L}{\partial\phi}-D\frac{\partial L}{\partial D\phi}-\bar{D}\frac{\partial L}{\partial \bar{D}\phi}\right)+
\delta t\left[D\left(\dot{\phi}\frac{\partial L}{\partial D\phi}\right)+\bar{D}\left(\dot{\phi}\frac{\partial L}{\partial \bar{D}\phi}\right)-\frac{d L}{dt}+\frac{\partial L}{\partial t}\right]\nonumber\\
&&-\xi\left[D\left(Q{\phi}\frac{\partial L}{\partial D\phi}\right)+\bar{D}\left(Q{\phi}\frac{\partial L}{\partial \bar{D}\phi}\right)\right]
+\bar\xi\left[D\left(\bar Q{\phi}\frac{\partial L}{\partial D\phi}\right)+\bar{D}\left(\bar Q{\phi}\frac{\partial L}{\partial \bar{D}\phi}\right)\right]\Bigg\}.\label{eqnsusy}
\end{eqnarray}
Further, taking into account (\ref{trafos}) and (\ref{dertotal1}), (\ref{dertotal2}), we get for supersymmetry transformations
\begin{equation}
\delta_{\xi} S=-\int_{t_{i}}^{t_{f}}dt\int d\theta d\bar\theta\left(\xi DL-\bar{\xi}\bar{D}L\right).
\end{equation}
Therefore, taking into account the equations of motion (\ref{ems}), the conservation laws of the conservative theory of lagrangian $L$ are
\begin{eqnarray}
\int_{t_{i}}^{t_{f}}dt\int d\theta d\bar\theta\left[D\left(\dot{\phi}\frac{\partial L}{\partial D\phi}\right)+\bar{D}\left(\dot{\phi}\frac{\partial L}{\partial \bar{D}\phi}\right)-\frac{d L}{dt}+\frac{\partial L}{\partial t}\right]&=&0,\label{nts1}\\
\int_{t_{i}}^{t_{f}}dt\int d\theta d\bar\theta\xi\left[D\left(Q{\phi}\frac{\partial L}{\partial D\phi}-L\right)+\bar{D}\left(Q{\phi}\frac{\partial L}{\partial \bar{D}\phi}\right)\right]&=&0,\label{nts2}\\
\int_{t_{i}}^{t_{f}}dt\int d\theta d\bar\theta\bar\xi\left[D\left(\bar Q{\phi}\frac{\partial L}{\partial D\phi}\right)+\bar{D}\left(\bar Q{\phi}\frac{\partial L}{\partial \bar{D}\phi}-L\right)\right]&=&0. \label{nts3}
\end{eqnarray}
From which, for a time independent Lagrangian, we obtain the expressions for the energy and the supersymmetric charges
\begin{eqnarray}
E&=&-i\left(\dot{\phi}\frac{\partial L}{\partial D\phi}\right)_\theta+i\left(\dot{\phi}\frac{\partial L}{\partial \bar D\phi}\right)_{\bar\theta}-L_{\theta\bar\theta},\label{nt1}\\
J&=&-\left(Q\phi\frac{\partial L}{\partial D\phi}-L\right)_\theta+\left(Q\phi\frac{\partial L}{\partial\bar D\phi}\right)_{\bar\theta},\label{nt2}\\
\bar J&=&-\left(\bar Q\phi\frac{\partial L}{\partial D\phi}\right)_\theta+\left(\bar Q\phi\frac{\partial L}{\partial\bar D\phi}-L\right)_{\bar\theta}.\label{nt3}
\end{eqnarray}
which satisfy $\dot E=0$, $\dot J=0$ and $\dot{\bar{J}}=0$. $L_{\theta\bar\theta}$ is the $\theta\bar\theta$ component of the superfield $L$.  Actually, these conservation laws can be given in a supersymmetric covariant way by the vanishing of the superfields inside the square brackets in (\ref{nts1}),  (\ref{nts2}) and  (\ref{nts3}).

Thus, for the nonconservative system, from  (\ref{eqsusy}) and (\ref{eqnsusy}) we get
\begin{equation}
\frac{dE}{dt}=\left(-\frac{\partial L}{\partial t}+\dot{\phi}F_K\right)_{\theta\bar{\theta}}, \qquad \frac{dJ}{dt}=\left[(Q\phi) F_K\right]_{\theta\bar{\theta}}\qquad {\rm and}
\qquad  \frac{d\bar J}{dt}=\left[ (\bar{Q}\phi) F_K\right]_{\theta\bar{\theta}}.
\end{equation}
If the conservative theory has additional, internal symmetries, say $\phi'(t,\theta,\bar{\theta})=\phi(t,\theta,\bar{\theta})+\alpha^a T_a\phi(t,\theta,\bar{\theta})$, which satisfy $[T_a,Q]=[T_a,\bar Q]=0$, then the transformation of the action under these transformations is
\begin{equation}
\delta_\alpha S=\int_{t_{i}}^{t_{f}}dt\int d\theta d\bar\theta\ 
\alpha^a\left[-D\left(T_a{\phi}\frac{\partial L}{\partial D\phi}\right)-\bar{D}\left(T_a{\phi}\frac{\partial L}{\partial \bar{D}\phi}\right)+T_a{\phi}\left(\frac{\partial L}{\partial\phi}-D\frac{\partial L}{\partial D\phi}-\bar{D}\frac{\partial L}{\partial \bar{D}\phi}\right)\right].
\end{equation} 
Therefore, the corresponding conserved current of the conservative theory is
\begin{equation}
J_a=D\left(T_a{\phi}\frac{\partial L}{\partial D\phi}\right)+\bar{D}\left(T_a{\phi}\frac{\partial L}{\partial \bar{D}\phi}\right),
\end{equation}
whose conservation law for the nonconservative theory as
\begin{equation}
\frac{dJ_a}{dt}=\left[(T_a\phi) F_K\right]_{\theta\bar{\theta}}.\label{otro}
\end{equation}
Unlike the case of time translations and supersymmetry transformations, a nontrivial nonconservative generalized potential $K$ can be invariant under the internal symmetry transformations of the conservative system, and (\ref{otro}) can be zero.
\subsection{Interacting oscillators}
Let us consider two interacting oscillators. As a first step we work out the supersymetric conservative action 
\begin{equation}
S=\int_{t_{i}}^{t_{f}}dt\int d\theta d\bar\theta\left[\frac{m}{2}\left(\bar D\phi D\phi+\omega\phi^{2}\right)-\lambda\phi\Phi+\frac{M}{2}\left(\bar D\Phi D\Phi+\Omega\Phi^{2}\right)\right],\label{accionsuper}
\end{equation}
where $\phi(t,\theta,\bar{\theta})=q(t)+\theta\psi(t)-\bar{\theta}\bar{\psi}(t)+\theta\bar{\theta}b(t)$ and $\Phi(t,\theta,\bar{\theta})=Q(t)+\theta\Psi(t)-\bar{\theta}\bar{\Psi}(t)+\theta\bar{\theta}B(t)$ (not confuse the variable $Q(t)$ with the supersymmetric charge $Q$). In components this action can be written as $S=S_{b}+S_{f}$, where
\begin{eqnarray}
S_{b}&=&\int_{t_{i}}^{t_{f}}\left[\frac{m}{2}\dot{q}^{2}+m\omega qb+\frac{m}{2}b^{2}-\lambda\left(qB+Qb\right)+\frac{M}{2}\dot{Q}^{2}+M\Omega QB+\frac{M}{2}B^{2}\right],\\
S_{f}&=&\int_{t_{i}}^{t_{f}}\left[-i\frac{m}{2}\left(\psi\dot{\bar{\psi}}+\bar\psi\dot\psi\right)+m\omega\psi\bar\psi-\lambda\left(\psi\bar\Psi+\Psi\bar\psi\right)
-i\frac{M}{2}\left(\Psi\dot{\bar{\Psi}}+\bar\Psi\dot\Psi\right)+M\Omega\Psi\bar\Psi\right].
\end{eqnarray}
The equations of motion of $Q$, $\Psi$, $\bar\Psi$ and $B$ are  
\begin{eqnarray}
\ddot Q+\left(\Omega^{2}+\frac{\lambda^{2}}{M^2}\right)Q-\frac{\lambda}{M}\left(\omega+\Omega\right)q=0, \label{e1}\\
i\dot\Psi+\Omega\Psi-\frac{\lambda}{M}\psi=0\label{e2},\\ -i\dot{\bar{\Psi}}+\Omega\bar\Psi-\frac{\lambda}{M}\bar\psi=0,\label{e3}\\
B+\Omega Q-\frac{\lambda}{M} q=0. \label{e4}
\end{eqnarray}
From them follow effective actions invariant under time reversal, as in previous section. These equations can be obtained from a superfield variation, such that $\delta\Phi(t_i)=\delta\Phi(t_f)=0$. The superfield equations of motion are
\begin{equation}
\frac{1}{2}\left[\bar D,D\right]\Phi-\Omega\Phi+\frac{\lambda}{M}\phi=0.\label{eqsuper}
\end{equation}
Its solution can be written as
\begin{equation}
\Phi(z)=\Phi^{(h)}(z)-\frac{\lambda}{M}\int dz'\Delta(z,z')\phi(z),\label{solsuper}
\end{equation}
where $z\equiv(t,\theta,\bar\theta)$ and $\int dz\equiv\int_{t_{i}}^{t_{f}}dt\int d\theta d\bar\theta$ and the propagator satisfies
\begin{equation}
\left(\frac{1}{2}\left[\bar D,D\right]-\Omega\right)\Delta(z,z')=\delta^{(3)}(z-z').\label{eqps}
\end{equation}
The Dirac function for anticommutative variables is defined as $\delta(\theta-\theta')=\theta-\theta'$ and satisfies $\int d\theta f(\theta)\delta(\theta-\theta')=f(\theta')$, hence $\delta^{(3)}(z-z')=\delta(t-t')\delta(\theta-\theta')\delta(\bar\theta-\bar\theta')=\delta(t-t')(\theta-\theta')(\bar\theta-\bar\theta')$. $\Delta(z,z')$ has a rather lengthy expansion in its anticommutative variables, but it can be written in a simpler form as $\Delta(z,z')=\Delta_{0}(t,t',\theta',\bar{\theta}')+\theta\Delta_{\theta}(t,t',\theta',\bar{\theta}')-\bar\theta\Delta_{\bar{\theta}}(t,t',\theta',\bar{\theta}')+\theta\bar\theta\Delta_{\theta\bar\theta}(t,t',\theta',\bar{\theta}')$, which substituted into (\ref{eqps}) and solved, component by component, leads to
\begin{equation}
\Delta_{ret}(z,z')=\left(1-\Omega\theta\bar\theta\right)\left(1-\Omega\theta'\bar{\theta}'\right)G_{ret}(t-t')+\theta\bar{\theta}'\Gamma_{ret}(t-t')+\bar\theta{\theta}'\bar\Gamma_{ret}(t-t')-\theta\bar\theta'\theta\bar{\theta}'\delta(t-t'),\label{delta}
\end{equation}
where $G_{ret}(t-t')=\pi/\Omega\, \sin\Omega(t-t')$ and $\Gamma_{ret}(t-t')$ is given in (\ref{gret}). Substituting (\ref{delta}) into (\ref{solsuper}) leads to the solutions of the equations (\ref{e1})-(\ref{e4}). Note that (\ref{delta}) does not depend on the difference $\theta-\theta'$.
In order to write the effective superfield action of (\ref{accionsuper}), we have 
\begin{eqnarray}
\bar D\Phi D\Phi&=&a\left[\bar D(\Phi D\Phi)-\Phi\bar DD\Phi\right]
-b\left[D(\Phi\bar D\Phi)-\Phi D\bar D\Phi\right]\\
&=&\frac{1}{2}\Phi\left\{(b-a)\{D,\bar D\}+(b+a)[D,\bar D]\right\}\Phi+a\bar D(\Phi D\Phi)-bD(\Phi\bar D\Phi)\\
&=&\frac{1}{2}\Phi[D,\bar D]\Phi+i\frac{a-b}{2}\frac{d}{dt}(\Phi^2)+a\bar D(\Phi D\Phi)-bD(\Phi\bar D\Phi),
\end{eqnarray}
where $a+b=1$. Thus, applying this result into the action (\ref{accionsuper}) and taking into account (\ref{eqsuper}) and (\ref{solsuper}), we get
\begin{equation}
S=\int dz\left\{\frac{m}{2}\left[\bar D\phi(z) D\phi(z)+\omega\phi^{2}(z)\right]-\frac{\lambda}{2}\phi(z)\Phi^{(h)}(z)+\frac{\lambda^{2}}{2M}\int dz'\phi(z)\Delta_{ret}(z,z')\phi(z')\right\}+{\rm B.T.},
\end{equation}
which is time reversal symmetric, hence conservative. The boundary terms of (\ref{cosce}), taking into account (\ref{dertotal1}) and (\ref{dertotal2}), are given by
\begin{equation}
\frac{m}{2}\int dz\left[a\bar D(\Phi D\Phi)-bD(\Phi\bar D\Phi)\right]=\frac{m}{2}\left.[i(a-b)(BQ+\Psi\bar\Psi)+Q\dot Q]\right\vert_{t_i}^{t_f},
\end{equation}
whose variation vanishes only if $\delta\dot Q(t_{i,f})=0$. 

The nonconservative action is $\Lambda(\Phi_{1},\Phi_{2},D\Phi_{1},\bar D\Phi_{2})=L(\Phi_{1},D\Phi_{1})-L(\Phi_{2},D\Phi_{2})$. This action can be written with the superfields $\phi_{\pm}=1/\sqrt{2}(\phi_{1}\pm\phi_{2})$ and $\Phi_{\pm}=1/\sqrt{2}(\Phi_{1}\pm\Phi_{2})$, giving
\begin{eqnarray}
S=\int dz&\bigg[&\frac{m}{2}\left(\bar D\phi_{+}D\phi_{-}+\bar D\phi_{-}D\phi_{+}\right)+m\omega\phi_{+}\phi_{-}\nonumber\\
&&-\lambda\left(\phi_{+}\Phi_{-}+\phi_{-}\Phi_{+}\right)+\frac{M}{2}\left(\bar D\Phi_{+}D\Phi_{-}+\bar D\Phi_{-}D\Phi_{+}\right)+M\Omega\Phi_{+}\Phi_{-}\bigg].
\end{eqnarray}
We are interested only on the equations of $\Phi_{\pm}$, which are the same as (\ref{eqsuper}), $\frac{1}{2}\left[\bar D,D\right]\Phi_{\pm}-\Omega\Phi_{\pm}+\frac{\lambda}{M}\phi_{\pm}=0$, with solution (\ref{solsuper}). In order to write the effective action, we have
\begin{eqnarray}
\bar D\Phi_{+}D\Phi_{-}=a\left[\bar D\left(\Phi_{+}D\Phi_{-}\right)-\Phi_{+}\bar DD\Phi_{-}\right]-b\left[D\left(\Phi_{-}\bar D\Phi_{+}\right)-\Phi_{-}D\bar D\Phi_{+}\right],\\
\bar D\Phi_{-}D\Phi_{+}=c\left[\bar D\left(\Phi_{-}D\Phi_{+}\right)-\Phi_{-}\bar DD\Phi_{+}\right]-d\left[D\left(\Phi_{+}\bar D\Phi_{-}\right)-\Phi_{+}D\bar D\Phi_{-}\right],
\end{eqnarray}
where $a+b=c+d=1$. From these equations, after some manipulations and taking into account the anticommutator $\{D,\bar D\}=-2id/dt$ and $a-d=c-b$, we get
\begin{eqnarray}
\bar D\Phi_{+}D\Phi_{-}+\bar D\Phi_{-}D\Phi_{+}&=&-\frac{1}{2}(a+d)\Phi_{+}[\bar D,D]\Phi_{-}-\frac{1}{2}(b+c)\Phi_{-}[\bar D,D]\Phi_{+}+i(a-d)\frac{d}{dt}\left(\Phi_{+}\Phi_{-}\right)\nonumber\\
&&+a\bar D\left(\Phi_{+}D\Phi_{-}\right)-bD\left(\Phi_{-}\bar D\Phi_{+}\right)+c\bar D\left(\Phi_{-}D\Phi_{+}\right)-d D\left(\Phi_{+}\bar D\Phi_{-}\right).\label{bts}
\end{eqnarray}
Thus, the effective action is given by
\begin{eqnarray}
S&=&\int dz\bigg\{\frac{m}{2}\left(\bar D\phi_{+}D\phi_{-}+\bar D\phi_{-}D\phi_{+}\right)+m\omega\phi_{+}\phi_{-}-\frac{b+c}{2}\lambda\phi_{-}\Phi_{+}^{(h)}\nonumber\\
&&+\frac{\lambda^{2}}{2M}\int dz\phi_{-}(z)\left[(b+c)\Delta_{ret}(z,z')+(a+d)\Delta_{ret}(z',z)\right]\phi_{+}(z')\bigg\}+{\rm B.T.}\label{accionefsi}
\end{eqnarray}
which is causal if $a+d=0$, and $b+c=2$. The boundary terms, which can be derived from (\ref{bts}), can be seen to vanish, which turns out from the last condition and the variational boundary conditions. Indeed
\begin{eqnarray}
{\rm B.T.}&=&\frac{M}{2}\big[i(a-d)(Q_{+}B_{-}+Q_{-}B_{+})-i(a-d)Q_{+}B_{-}+i(b-c)Q_{-}B_{+}-i(a-b)\Psi_{-}\bar\Psi_{+}\nonumber\\&&-i(c-d)\Psi_{+}\bar\Psi_{-}
+(a+d)Q_{+}\dot Q_{-}+(b+c)Q_{-}\dot Q_{+}\big]\Big\vert_{t_i}^{t_f}=(a+d)Q_{+}(t_{i})\dot Q_{-}(t_{i})=0.
\end{eqnarray}
In components, after elimination of the auxiliary fields $b(t)$ and $B(t)$, we obtain from (\ref{accionefsi})
\begin{eqnarray}
S_{b,{\rm eff}}&=&\int_{t_{i}}^{t_{f}}dt\left[m\dot{q}_{+}\dot{q}_{-}-m\left(\omega^{2}+\frac{\lambda^{2}}{m}\right){q}_{+}{q}_{-}+\lambda(\omega+\Omega) q_{-}Q_{+}^{(h)}+\frac{\lambda^{2}}{M}(\omega+\Omega)^{2}\int_{t_{i}}^{t_{f}}dt'q_{-}(t)G_{ret}(t-t')q_{+}(t')\right],\\
S_{f,{\rm eff}}&=&\int_{t_{i}}^{t_{f}}dt\left[-i\frac{m}{2}\left(\psi_{+}{\dot{\bar\psi}}_{-}+{\bar\psi}_{-}{\dot{\psi}}_{+}\right)+m\omega\psi_{+}{\bar\psi}_{-}-\lambda\psi_{-}{\bar\Psi}_{+}^{(h)}\right]\nonumber\\
&-&\frac{\lambda^{2}}{M}\int_{t_{i}}^{t_{f}}dt\int_{t_{i}}^{t_{f}}dt'\left[\psi_{-}(t)\Gamma_{ret}(t-t'){\bar\psi}_{+}(t')-{\bar\psi}_{-}(t)\bar\Gamma_{ret}(t-t'){\psi}_{+}(t')\right]
+{\rm c.c.},
\end{eqnarray}
where $G_{ret}(t-t')=1/\sqrt{\Omega^{2}+\frac{\lambda^{2}}{M^{2}}}\ \sin\sqrt{\Omega^{2}+\frac{\lambda^{2}}{M^{2}}}(t-t')$ and $\Gamma_{ret}(t-t')=-i\theta(t-t')e^{i\Omega(t-t')}$.

The generalization for the interaction of more than one oscillator $Q$ can be done straightforwardly following \cite{galley1}.
\subsection{Damped supersymmetric oscillator}
As a last example we consider the damped oscillator, and compute its conservation laws. The conservative lagrangian is the one of the oscillator $L(\phi,D\phi,\bar D\phi)=\frac{m}{2}\left(\bar D\phi D\phi+\omega\phi^{2}\right)$ and $K(\phi_1,D\phi_1,\bar D\phi_1,\phi_2,D\phi_2,\bar D\phi_2)=i\tilde c/2(\bar D\phi_1 D\phi_2-\bar D\phi_2 D\phi_1)$. The action of the nonconservative lagrangian $\Lambda=L(\phi_1)-L(\phi_2)+K(\phi_1,\phi_2)$ is in components
\begin{eqnarray}
S=\int_{t_{i}}^{t_{f}}dt&\biggl[&\frac{m}{2}\left(\dot q_1^2-\dot q_2^2\right)-i\frac{m}{2}\left(\psi_1\dot{\bar{\psi}}_1+\bar\psi_1\dot{\psi}_1-\psi_2\dot{\bar{\psi}}_2-\bar\psi_2\dot{\psi}_2\right)+m\omega\left(q_1b_1+\psi_1\bar{\psi}_1-q_2b_2-\psi_2\bar{\psi}_2\right)\nonumber\\
&&+\frac{m}{2}\left(b_1^2-b_2^2\right)-\tilde c\left(b_1\dot{q}_2-b_2\dot{q}_1\right)+\frac{\tilde c}{2}\left(\bar\psi_2\dot{\psi}_1-\bar\psi_1\dot{\psi}_2+\psi_1\dot{\bar{\psi}}_2-\psi_2\dot{\bar{\psi}}1\right)\biggl].
\end{eqnarray}
As usual, the auxiliary fields can be eliminated by their equations of motion $b_1+\omega q_1-c/m\dot{q}_2=0$ and $b_2+\omega q_2-c/m\dot{q}_1=0$, with the resulting action
\begin{eqnarray}
S=\int_{t_{i}}^{t_{f}}dt&\biggl[&\frac{m}{2}\left(1+\frac{\tilde c^2}{m^2}\right)\left(\dot q_1^2-\dot q_2^2\right)-\frac{m}{2}\left(q_1^2-q_2^2\right)-i\frac{m}{2}\left(\psi_1\dot{\bar{\psi}}_1+\bar\psi_1\dot{\psi}_1-\psi_2\dot{\bar{\psi}}_2-\bar\psi_2\dot{\psi}_2\right)\nonumber\\
&&+m\omega\left(\psi_1\bar{\psi}_1-\psi_2\bar{\psi}_2\right)+\omega \tilde c\left(q_1\dot{q}_2-q_2\dot{q}_1\right)+\frac{\tilde c}{2}\left(\bar\psi_2\dot{\psi}_1-\bar\psi_1\dot{\psi}_2+\psi_1\dot{\bar{\psi}}_2-\psi_2\dot{\bar{\psi}}1\right)\biggl].
\end{eqnarray}
The equations of motion in the physical limit are (\ref{eqfnc1}),  (\ref{eqfnc2}) and
\begin{eqnarray}
\ddot q-2\tilde\omega\frac{\tilde c}{m}\dot{q}+\tilde\omega^2\left(1+\frac{\tilde c^2}{m^2}\right) q=0.\label{eqbnc}
\end{eqnarray}
The energy and the supersymmetric charges of the conservative system are
\begin{eqnarray}
E&=&\frac{m}{2}\left(\dot{q}^2+\omega^2q^2\right)-\omega\psi\bar\psi,\\
J&=&\left(i\dot q+\omega q\right)\psi,\\
\bar J&=&-\left(i\dot q-\omega q\right)\bar\psi.
\end{eqnarray}
Their time variations are
\begin{eqnarray}
\frac{dE}{dt}&=&2\tilde{c}\left(-\omega\dot{q}^2+\dot{\psi}\dot{\bar{\psi}}\right)=2\tilde{c}\tilde{\omega}\left[-\left(1+\frac{\tilde{c}^2}{m^2}\right)\dot{q}^2+\omega\psi\bar\psi\right],\\
\frac{dJ}{dt}&=&\tilde{c}\left[2i\dot{q}\dot{\psi}+\omega\left(q\dot{\psi}-\dot{q}\psi\right)\right]=\frac{1}{\tilde{\omega}}\left[\ddot{q}-3i\tilde{\omega}\left(1-i\frac{\tilde{\omega}}{m}\right)\dot{q}\right]\dot{\psi},
\end{eqnarray}
and the complex conjugate of the second equation. For the r.h.s. of both equations, the equations of motion  (\ref{eqfnc1}) and (\ref{eqbnc}) have been used.
\section{Conclusions}
We have generalized the variational formalism of \cite{galley1} for nonconservative systems for fermionic and supersymmetric systems. Consistency with the first order equations of motion of fermionic variables, requires that the boundary conditions are slightly modified, with no further consequences. Otherwise the generalization is straightforward in the superfield formalism of supersymmetry. Similar to the case of time translational symmetry, for supersymmetric theories we maintain the supersymmetric structure for the nonconservative generalized potential, which is written in terms of superfields. The Noether theorem is evaluated and as expected, the supersymmetric charges are not conserved, unlike the case of possible internal symmetries, which could be conserved. In the case of coupled oscillators, the derivation of the effective lagrangian involves boundary terms which can be problematic in the bosonic version, but which can be consistently neglected in the fermionic and supersymmetric versions. It would be interesting to generalize this approach for local supersymmetry, and explore consequences in supersymmetric quantum mechanics, as well as for field theory, in particular for supergravity.

\vskip 1truecm
\centerline{\bf Acknowledgements}
We thank VIEP-BUAP and PIFI-SEP for the support.


\end{document}